\documentclass[%
reprint,
  amsmath,amssymb,
  aps,
prl,
]{revtex4-1}

\usepackage{graphicx}
\usepackage{dcolumn}
\usepackage{bm}
\usepackage{color}
\usepackage{ulem}
\usepackage{natbib}

\usepackage{blindtext}
\usepackage{subfiles}

\begin{document}

\title{Anomalous relaxation of coarsening foams with viscoelastic continuous phase}

\author{Chiara Guidolin$^{1,2*}$, Emmanuelle Rio$^1$, Roberto Cerbino$^3$, Anniina Salonen$^{1*}$,  and Fabio Giavazzi$^{2*}$}
\affiliation{$^1$Université Paris-Saclay, CNRS, Laboratoire de Physique des Solides, Orsay, France.\\
$^2$Department of Medical Biotechnology and Translational Medicine, University of Milan, Segrate, Italy.\\
$^3$Faculty of Physics, University of Vienna, Vienna, Austria.\\
$^*$chiara.guidolin@unimi.it,
fabio.giavazzi@unimi.it,
anniina.salonen@universite-paris-saclay.fr}

\begin{abstract}
We investigate the ultraslow structural relaxation of ageing foams with rheologically-tunable continuous phases.
We probe the bubble dynamics associated with pressure-driven foam coarsening using differential dynamic microscopy, which allows characterizing the sample dynamics in the reciprocal space with imaging experiments.
Similar to other out-of-equilibrium jammed soft systems, these foams exhibit compressed exponential relaxations, with a ballistic-like linear dependency of the relaxation rate on the scattering wavevector.
By tuning the rheology of the continuous phase we observe changes in the relaxation shape, where stiffer matrices yield larger compressing exponents.
Our results corroborate recent real-space observations obtained with bubble tracking, providing a comprehensive overview of structural relaxation in these complex systems, both in direct and reciprocal space.
\end{abstract}

\maketitle

\noindent

\section{Introduction}

Jammed soft materials, such as dense emulsions, pastes, foams, and colloidal gels, are ubiquitous in daily life and have numerous industrial applications.
Despite their widespread use, these materials often exist out of equilibrium and undergo complex aging processes\cite{cloitre2000rheological, marze2009oscillatory}, aspects that are poorly understood and leave a significant gap in knowledge regarding their dynamic properties as they slowly evolve towards equilibrium.

A particularly intriguing feature of these jammed systems is their similarity to molecular glasses \cite{donth2001glassbook}.
Indeed, the tight packing and/or the presence of strong interactions between the components markedly reduce their mobility, lending the system a glassy dynamical behavior, consisting in very slow relaxations, non-exponential response or correlation functions, history-dependent dynamics, and dynamical heterogeneity \cite{Cipelletti2005slow}.

However, instead of the stretched exponential relaxations and diffusive dispersion relations typically found in hard glasses, dynamic light scattering experiments on jammed soft systems often give intermediate scattering functions displaying compressed exponential decays, with an anomalous linear dependence of the relaxation rate on the scattering wavevector, reminiscent of ballistic motion \cite{ramos2001ultraslow}.
This unusual dynamics has been ascribed to the relaxation of internal stresses.
It was proposed that randomly distributed dipolar stress sources generate displacement fields that lead to directionally persistent displacements characterised by a power-law tailed probability distribution \cite{bouchaud2001MODEL}.
The same compressed exponential, ballistic-like dynamics has been observed in a large variety of disordered, jammed, soft materials, including colloidal fractal gels, concentrated emulsions, micellar polycrystals, and lamellar gels, suggesting the generality of this behavior \cite{cipelletti2003universal}.

Stress-driven dynamics has been widely explored in coarsening aqueous foams \cite{durian1990dynamics,durian1991multiple,Durian1991_Gillette,cohen2001bubble,Gittings2008,Sessoms2010,Giavazzi2021}, where the source of stress imbalances can be traced back to the continuous bubble growth.
The pressure-driven coarsening process keeps altering the stress configuration of the system, leading to locally imbalanced stresses that trigger neighbour-switching bubble rearrangements.
A combined reciprocal and direct space analysis has recently shown how foam dynamics is governed by intermittent bubble displacements exhibiting a persistent direction up to the bubble size length scale \cite{Giavazzi2021}.
This length scale introduces a cut-off in the probability distribution function of bubble displacements that otherwise exhibits power-law scaling.
The existence of such cut-off length leads to a loss of linearity and the identification of a steeper second low-$q$ regime for the $q$-dependence of the foam relaxation rate \cite{Giavazzi2021}.

While all this is known for aqueous foams, what happens in more complex systems is much less understood, even though in many practical applications foams are themselves made from fluids with non trivial rheology, like gels, emulsions, or pastes.
Continuous phase rheology can modify the internal bubble dynamics during coarsening, by slowing down \cite{LeMerrer2012PRL, LeMerrer2013PRE} or even preventing \cite{guidolin_tracking} mutual bubble rearrangements, eventually impacting the overall foam morphology \cite{Guidolin2023}.
Stress relaxation thus plays a central role in the long term foam structure evolution.
A deep understanding of bubble dynamics during foam ageing would thus allow for a finer control of foam stability and mechanical properties which is of fundamental importance for many foam applications.

In this study, we examine how the rheological properties of the continuous phase influence the internal structural relaxation of foams during their destabilization.
We focus on bubble dynamics within coarsening foams made of concentrated oil-in-water emulsions.
At oil volume fractions beyond random close packing, these dense biliquid dispersions exhibit viscoelasticity, where both the storage modulus and yield stress increase with $\phi$\cite{Mason1995, Mason1996}.
Consequently, the rheological properties of the continuous phase of the foam can be adjusted by altering the emulsion oil fraction.

Recent findings suggest that higher emulsion yield stresses significantly slow down bubble movement during coarsening by accumulating elastic stresses in the matrix, which decouple bubble growth from dynamics \cite{guidolin_tracking}.
Nonetheless, the correlation between stress-induced displacements and their signatures in reciprocal space remains largely unexplored.
Filling this gap is crucial given that neutron and x-ray experiments on soft jammed materials cannot be easily performed in direct space, and one relies often on reciprocal scale information.

We use differential dynamic microscopy (DDM)\cite{CerbinoTrappe2008} to obtain reciprocal space information about the foam relaxation, demonstrating that increasing the emulsion oil fraction modifies the shape of the intermediate scattering function, resulting in more pronounced compressed exponential decays. Furthermore, the lack of bubble rearrangements at high emulsion yield stress \cite{guidolin_tracking} is mirrored in reciprocal space by alterations in the $q$-dependence of the foam relaxation rate.
The typically observed superlinear low-$q$ regime in coarsening aqueous foams gradually shifts towards a more linear dispersion relation with increased matrix stiffness, indicative of ultraslow ballistic-like bubble motion.

Our findings elucidate the impact of continuous phase rheology on foam relaxation, as observed in reciprocal space.
Moreover, this work provides a detailed analysis of structural relaxation in complex foam systems, representing a foundational study of the interplay between reciprocal and direct space dynamic features potentially extendable to a broad range of soft glassy systems.

\section{Materials and methods}

\subsection*{Sample preparation and imaging}

Our samples are foams made of concentrated oil-in-water emulsions.
Emulsions are first generated by mechanically mixing the oil and the aqueous phase with the double-syringe technique.
A syringe (Codan Medical, 60 mL) is partially filled with a volume $V_\text{oil}$ of rapeseed oil (from Brassica Rapa, Sigma Aldrich), while a second one is partly filled with a volume $V_\text{aq}$ of surfactant solution (Sodium Dodecyl Sulphate, Sigma Aldrich) at 30 g/L in deionised water.
The volumes $V_\text{oil}$ and $V_\text{aq}$ are chosen according to the desired emulsion oil volume fraction, defined as $\phi = V_\text{oil}/(V_\text{oil}+V_\text{aq}$).
The two syringes are then connected with a double luer lock and the syringe plungers are pushed 30 times back and forth.
The syringe inlets, having an inner diameter of 2 mm, act as constrictions in the flow of the mixture, breaking the oil phase into micrometric droplets.
The final result is a stable emulsion with droplet volume-weighted diameter distribution centered around 5 $\mu$m, as assessed with laser diffraction granulometry (Mastersizer 3000E, Malvern Panalytical) after emulsion generation.
The surface tension of the air/emulsion interfaces is $\sim$ 30 mN/m, as measured using a pendant drop tensiometer (Tracker, Teclis, France).
The oil fractions investigated are all above the random close packing fraction and range between 65\% and 80\%.
We thus vary the elastic modulus of the foam continuous phase between 30 and 340 Pa, and its yield stress from 0.5 to 20 Pa \cite{Guidolin2023, guidolin_tracking}.

Emulsions are foamed with the aid of a planetary kitchen mixer (Kenwood MultiOne, 1000 W).
Freshly-generated emulsion is poured in the vessel and the mixer is operated at increasing speed until the sample volume has increased tenfold.
The final foam liquid fraction is measured by weight after generation, and its value oscillates around 11\%.
The foaming process does not change the drop size distribution of the emulsions, which are stable throughout the experiment, as checked with laser diffraction granulometry.
The surfactant concentration in the aqueous phase ensures complete surface coverage of both oil droplets and bubbles at the oil and gas volume fractions considered.

After generation, the foamed emulsion is gently sandwiched between two square glass plates (edge 20 cm) separated by a rubber joint of thickness 10 mm.
The spacing is much larger than the typical bubble size ($R\sim 10^{-1}$ mm), so that the cell contains several layers of bubbles and the foam sample can be safely considered three dimensional.
Nevertheless, what we probe is the dynamics of the bubble layer at the interface between the foam and the top glass plate, which is actually representative of the bulk coarsening \cite{pasquet2023aqueous,pasquet2023coarsening}.
Foam ageing is indeed monitored from the top by taking image stacks with a camera (Basler acA3800-14um, equipped with a Tamron lens 16 mm F/1.4), while a square array of LED lights provides uniform illumination from above.
The effective frame scale is 24.2 pixel/mm.
Images are acquired every 5 seconds ($\phi$ = 65\%, 70\%, 75\%) or 15 seconds ($\phi$=80\%) to ensure a proper time resolution for DDM analysis.
The emulsion yield stress allows delaying gravitational drainage \cite{Goyon2010_PRL_drainage}, so that foam coarsening can be probed in absence of significant gravity-induced vertical gradients in the liquid fraction.

\subsection*{Reciprocal space analysis}

Bubble dynamics is analysed in the reciprocal space using the DDM protocol \cite{CerbinoTrappe2008,giavazzi2009scattering}.
Raw frames are first cropped around a square region of interest (2048 x 2048 pixels) in which the illumination is rather uniform.
We correct for residual uneven illumination by dividing each frame by a background image, obtained by applying a Gaussian filter having a standard deviation of 20 mm to the first frame of the image stack.
The width of the Gaussian filter is chosen to be much larger than the typical bubble size but smaller than the extension of the intensity gradient due to uneven illumination.

Since foams are evolving over time, foam dynamics is studied in quasi-stationary conditions \cite{Giavazzi2021} by restricting the DDM analysis on three non-overlapping image sub-sequences, centered at different foam ages $t^*=$ 1800, 2700, and 3600 seconds respectively, and covering a $t^*/4$ time window in which the mean bubble size grows less than 15\% \cite{guidolin_tracking}.

Each sub-sequence is analysed separately with a custom MATLAB script as follows.
The difference between two background-corrected frames acquired at times $t$ and $t+\Delta t$, namely $\Delta I(x,t,\Delta t) = I(x,t+\Delta t) - I(x, t)$, is first calculated for different log-spaced $\Delta t$.
A 2D fast Fourier transform algorithm is then applied to $\Delta I(x,t,\Delta t)$, and the spatial Fourier power spectra obtained for the same lag time $\Delta t$ but different reference times $t$ inside the sequence are then averaged.
This way we obtain the image structure function $d(\textbf{q},\Delta t)$ which captures the sample dynamics as a function of the 2D scattering wavevector $\mathbf{q}$ and of the lag time $\Delta t$.
We then exploit the isotropy of the sample to calculate the azimuthal average of $d(\textbf{q}, \Delta t)$, which provides $d(q, \Delta t)$ as a function of the radial wavewector $q = \sqrt{q_x^2 + q_y^2}$.
For each $q$, the image structure function $d(q, \Delta t)$ is typically a monotonically increasing function of the delay time $\Delta t$, and is linked to the real part of the intermediate scattering function $f(q, \Delta t)$ by the relation $d(q,\Delta t)=A(q)[1-f(q,\Delta t)]+B(q)$, where the term $B(q)$ accounts for the camera noise, and the term $A(q)$ is the static amplitude.

By fitting a suitable model to $d(q, \Delta t)$, one can thus extract $f(q, \Delta t)$, whose decay encloses the information on the sample dynamics at a length scale $2\pi /q$.
A function of the kind $d(q,\Delta t)=A(q)[1-f(q,\Delta t)]+B(q)$, where $f(q,\Delta t)$ is a compressed exponential function $f(q,\Delta t)=\exp[-(\Gamma(q) \Delta t)^{\alpha(q)}]$, is then fitted to the image structure function data points for each image sub-sequence.
We could fit $d(q,\Delta t)$ to extract both $A(q)$ and $f(q,\Delta t)$.
However, since the analysis is restricted on a limited time interval for quasi-stationarity, a full relaxation of $f(q, \Delta t)$ is not observed at all wavevectors $q$.
We thus estimate the amplitude from the time-averaged power spectrum of the single images as $A(q) \simeq 2 \langle |\hat{I}(q)|^2 \rangle - B(q)$ \cite{cerbino2017dark, giavazzi2018tracking}.
This approximation is justified as the optical signal due to the foam is much higher than any other contribution coming from stray light or dirt on the optical components.
We first fit $d(q, \Delta t)$ normalising with the static amplitude $A(q)$ but leaving the compressing exponent $\alpha$ as a free parameter, to check its dependence on $q$.
The obtained compressing exponents $\alpha(q)$ are almost constant, displaying a weak $q$-dependence only for the larger $\phi$ values (Supplementary Information, Fig. S1).
We thus evaluate the average compressing exponent $\alpha = \langle \alpha(q) \rangle$, and perform a more constrained fit of $d(q,\Delta t)$ fixing the compressing exponent to be equal to its mean value $\alpha$, in order to extract the relaxation rate $\Gamma(q)$ from the decay of the intermediate scattering function $f(q,\Delta t)$.
In order to compare samples having different values of $\alpha$, we use the average relaxation rate $\langle\Gamma(q)\rangle = \Gamma(q) \cdot \alpha/\pmb{\Gamma}(1/\alpha)$, where $\pmb{\Gamma}$ denotes the gamma function.

\subsection*{Direct space analysis}

Real-space image processing is performed as described in previous work \cite{guidolin_tracking}.
Briefly, raw foam images are processed with custom MATLAB scripts as follows.
After background correction, the same images used for DDM analysis are segmented with an adaptive thresholding and then skeletonised with a water-shed algorithm.
The typical number of detected bubbles is always larger than $10^3$ at each foam age, ensuring good statistics.
The size of each single bubble is estimated from the foam skeleton as the radius $R=\sqrt{A/\pi}$, where $A$ is the area of the polygonal cell outlining the bubble boundaries.
The average bubble size at a given foam age $t^*$ is then calculated as the ensemble mean bubble radius of the corresponding frame, namely $R^*=\langle R\rangle$.

Bubble dynamics is then characterised in the real space at different foam ages $t^*$ by performing bubble tracking with TrackMate \cite{TrackMate_1, TrackMate_2} on the segmented foam images within the time window considered for quasi-stationarity.
Only bubbles whose trajectories cover the whole time window are kept for the analysis, whereas small bubbles that disappear for coarsening or blink for misdetection are discarded.
The latter however correspond to less than 20\% of the total number of detected bubbles.

\begin{figure*}[htb]
    \centering
    \includegraphics[width=.9\textwidth]{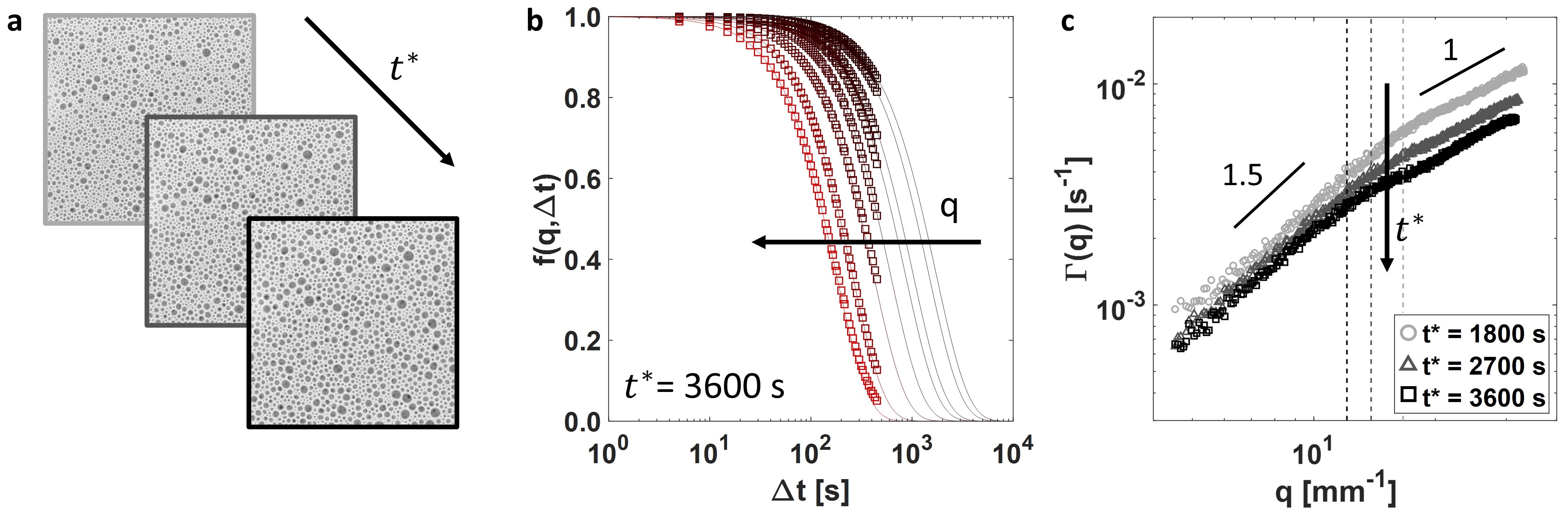}
    \caption{\textbf{DDM on coarsening foamed emulsions.} (a) Representative images of foamed emulsion at $\phi$ = 65\% at different foam ages $t^*$ = 1800, 2700, and 3600 seconds. Bubbles appear as dark spots embedded in the whitish emulsion forming the foam continuous phase.
    (b) Representative curves of the intermediate scattering function $f(q,\Delta t)$ obtained at different wavevectors $q$ for the sample $\phi$=65\% at $t^*=$ 3600 s.
    (c) Relaxation rate $\Gamma(q)$ obtained for the same sample at increasing foam ages $t^*$ equal to 1800 s, 2700 s, and 3600 s. The vertical dashed lines highlight the wavevectors $q^*=2\pi/R^*$, corresponding to characteristic bubble sizes.}
\label{fig:FIGURE_1}
\end{figure*}

\section{Results and discussion}

Foams age via gas diffusion from smaller bubbles to larger ones driven by differences in their Laplace pressure. \cite{VonNeumann1952}
This pressure-driven coarsening process makes the mean bubble size grow over time until complete phase separation. \cite{Mullins1986}Bubble size variations continuously generate stress imbalances within the foam, which eventually relax through local bubble rearrangements involving an exchange of neighbours.\cite{Cantat2013book}

The systems under study are foams made of air bubbles tightly embedded in concentrated oil-in-water emulsions.
The presence of viscoelastic emulsion between the bubbles has been observed to strongly affect the long-term coarsening evolution\cite{Guidolin2023} as well as the associated internal bubble dynamics \cite{guidolin_tracking}. Previous work revealed that, at low emulsion oil fractions, such systems behave like aqueous foams: bubbles freely rearrange during coarsening, so that they move persistently in one direction only up to a critical length scale of the order of the bubble size.
By contrast, the increase of the emulsion yield stress strongly hinders mutual bubble displacements, as the emulsion can bear stresses coming from bubble size variations up to its yielding point.
The lack of bubble rearrangements prevents sudden changes of direction in the bubble trajectories: bubbles keep moving persistently, deforming the material without relaxing the accumulated stress via neighbour-switching events.

\subsection*{DDM captures the coarsening-induced relaxation dynamics}

Here, we probe how the observed change in the bubble dynamics is reflected in the reciprocal space.
We thus perform differential dynamic microscopy on the same foam systems, varying the emulsion oil fraction $\phi$ between 65\% and 80\%. For each sample, we probe the bubble dynamics at three different foam ages $t^*$ corresponding to 1800, 2700, and 3600 seconds respectively.
The typical foam appearance is exemplified in Fig. \ref{fig:FIGURE_1}(a) for $\phi$ = 65\%.
At each foam age, we compute the intermediate scattering function $f(q,\Delta t)$.
A few representative examples, calculated at different wavevectors $q$, are shown in Fig. \ref{fig:FIGURE_1}(b) for the sample $\phi$=65\% at age $t^*$ = 3600 s, but extended results can be found in Supplementary Information (Fig. S2-S4).
The intermediate scattering function $f(q,\Delta t)$ shifts towards smaller $\Delta t$ with increasing $q$, reflecting the slower relaxation of larger length scales.

As we restrict the determination of $f(q,\Delta t)$ to a time window over which the foam can be considered quasi-stationary, the accessible data range is limited and a full decay is only obtained for large $q$-values.
Despite this limitation, we can access the decay rate of $f(q,\Delta t)$ by fitting these curves with a model function.
We however restrict our analysis to those wavevectors for which at least the first 10\% of the decay is visible.

The curves are well fitted by a compressed exponential function of the kind $f(q,\Delta t)=\exp[-(\Gamma(q) \Delta t)^\alpha]$, as in shaving foam \cite{Giavazzi2021}, with a compressing exponent $\alpha \simeq$ 1.35. The corresponding foam relaxation rate $\Gamma(q)$ obtained for the sample $\phi=65$\% at different foam ages is shown in Fig. \ref{fig:FIGURE_1}(c).
The curves shift downwards with time, mirroring the slowdown of bubble dynamics as the foam coarsens.
At each foam age, two distinct dynamic regimes are observed, as reflected by two different slopes in $\Gamma(q)$: while at high $q$ the dispersion relation is linear, at low $q$ the relaxation rate exhibits a steeper $q$-dependence.
A similar behavior was observed for shaving foam, where the dispersion relation was found to turn from a linear into a power-law scaling $\Gamma(q)\sim q^\beta$ (with $\beta>1$) for wavevectors corresponding to length scales above the typical bubble size \cite{Giavazzi2021}.
In shaving foams, the high-$q$ linear scaling has be ascribed to the ballistic-like bubble motion under the strain field induced by stress imbalances within the foam.
The occurrence of bubble rearrangements suddenly changes the local stress configuration, causing a loss of directional persistence in real space that is mirrored by the change of slope into a stronger $q$-dependence of $\Gamma(q)$ at low $q$.
This picture holds also for the foamed emulsion at $\phi$ = 65\%, as real-space bubble motion resembles the one of shaving foam \cite{guidolin_tracking}.
Moreover, the crossover between the two regimes occurs at a wavevector $q^* = 2\pi/R^*$, corresponding to the typical bubble size $R^*$, and shifts towards smaller $q$ over time, consistently with the gradual increase of $R^*$ as the foam ages.
At low $\phi$, coarsening foamed emulsions thus share the same dynamic features of classic aqueous foams in both real and reciprocal space.

\subsection*{The relaxation dynamics depends on matrix elasticity}

\begin{figure*}[htb]
    \centering
    \includegraphics[width=.9\textwidth]{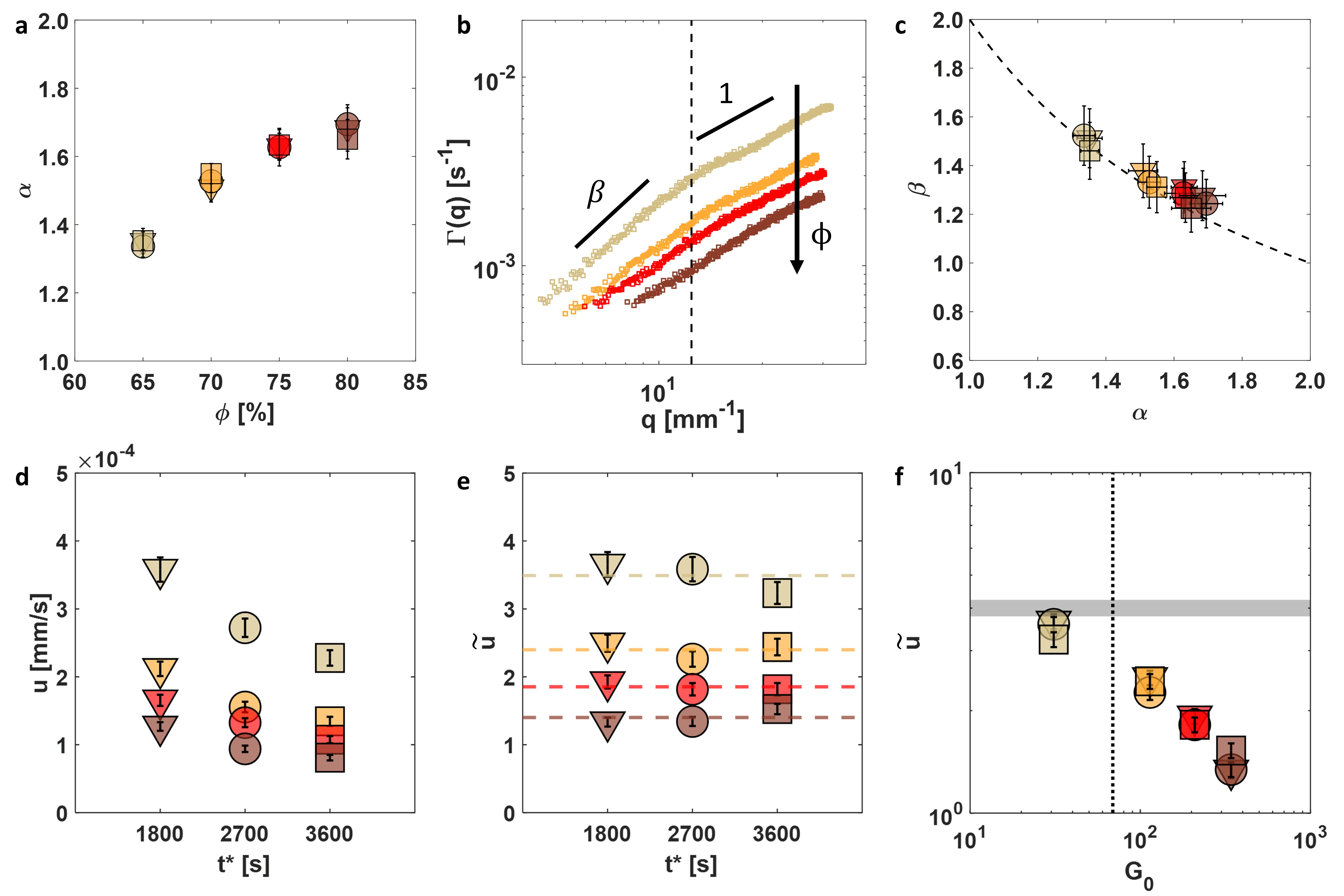}
    \caption{\textbf{Foam relaxation at different continuous phase elasticity.} (a) Compressing exponent $\alpha$ for samples at different oil fractions $\phi$ and different foam ages $t^*=$ 1800 s (triangles), 2700 s (circles), and 3600 s (squares).
    (b) Relaxation rates $\Gamma(q)$ obtained for different $\phi$ at fixed foam age $t^*=$ 3600 s.
    The slope $\beta$ highlights the super-linear scaling of the low-$q$ regime ($q<q^*$).
    The vertical dashed line marks the crossover wavevector $q^*=2\pi/R^*$.
    (c) Comparison between exponents $\beta$ and $\alpha$.
    The dashed line corresponds to the expected relation $\beta = 2/\alpha$.
    (d) Velocity $u$ obtained from a linear fit of the kind $\Gamma(q) = u q$ in the high-$q$ regime ($q>q^*$) for different foamed emulsions at different foam ages $t^*$.
    (e) Same velocities after normalisation with the coarsening rate, namely $\Tilde{u}=u/\dot R$.
    (f) Variation of the normalised velocity $\Tilde{u}$ with emulsion elasticity.
    The vertical dashed line marks the critical elastic modulus $G_0^* \sim \gamma/R$.
    The horizontal bar marks the reference value of $\Tilde{u}$ measured for shaving foam \cite{Giavazzi2021}.
    }
\label{fig:FIGURE_2}
\end{figure*}

We now repeat the same analysis for samples at different oil fractions.
At each foam age, the shape of the intermediate scattering function becomes more compressed with stiffer emulsion, as reflected by the increase of the compressing exponent $\alpha$ with increasing $\phi$ shown in Fig. \ref{fig:FIGURE_2}(a). The corresponding relaxation rates are compared in Fig. \ref{fig:FIGURE_2}(b) for the same foam age $t^*=3600$ s.
The comparison reveals that $\Gamma(q)$ decreases with increasing oil fraction, reflecting a slowing down of the coarsening-induced dynamics.
We stress that this slowing down is not due to any difference in the coarsening rate $\dot{R}=\text{d}R/\text{d}t$, as the coarsening kinetics does not change between the samples \cite{guidolin_tracking}.
The reduction can thus be entirely ascribed to the change in the continuous phase rheology.

Remarkably, the $q$-dependence of $\Gamma(q)$ also changes with $\phi$.
We can see that, while at low $\phi$ one can clearly distinguish two distinct dynamic regimes, such distinction is gradually lost as $\phi$ is increased up to 80\%, where $\Gamma(q)$ almost exhibits a single ballistic-like scaling over the whole range of accessible wavevectors.
A power-law fit of $\Gamma(q)$ in the low-$q$ regime (namely for $q<2\pi/R^*$) indeed reveals that the exponent $\beta$ decreases with increasing $\phi$ heading towards a linear scaling.
We recall that the low-$q$ regime is directly linked to the presence of the cut-off length: if the probed length scale exceeds the largest bubble displacement, the relaxation rate transitions to a different dispersion relation $\Gamma(q) \sim q^\beta$, with $\beta = 2/\alpha$ \cite{Giavazzi2021}, as supported also by our results shown in Fig. \ref{fig:FIGURE_2}(c).

Thus, in foamed emulsions, compressed exponential relaxations are associated with a high-$q$ linear dispersion relation combined with a steeper low-$q$ regime whose $q$-dependence changes with $\phi$ according to the variation of the compressing exponent $\alpha$.

Let us now focus on the high-$q$ linear scaling of $\Gamma(q)$.
On length scales below the bubble size, namely at $q>2\pi/R^*$, the relaxation rate can be expressed as $\Gamma(q) = u q$, with $u$ representing a characteristic relaxation velocity.
For a given foam sample, $\Gamma(q)$ decreases with increasing foam age, reflecting the slowing down of the coarsening kinetics over time, and so does the characteristic velocity $u$, as shown in Fig. \ref{fig:FIGURE_2}(d).
For a coarsening aqueous foam, dispersion relations $\Gamma(q)$ obtained at different foam ages collapse onto a single master curve after rescaling $q$ with the bubble size $R$ and $\Gamma$ with the strain rate $\dot{R}/R$ associated with coarsening \cite{Giavazzi2021}.
This means that the characteristic velocity $u$ normalised with the coarsening rate $\dot{R}=\text{d}R/\text{d}t$, namely $\Tilde{u} = u / \dot{R}$, remains constant over time over a large range of bubble sizes.
In our samples, the normalised velocity $\Tilde{u}$ no longer displays a significant variation over time, as shown in Fig. \ref{fig:FIGURE_2}(e), and can thus be considered constant at least within the range of bubble sizes investigated (as $t^*$ is doubled from 1800 s to 3600 s, we register a $\sim$ 40\% increase of $R^*$ from 0.37 to 0.51 mm).

\begin{figure*}[htbp]
    \centering
    \includegraphics[width=.9\textwidth]{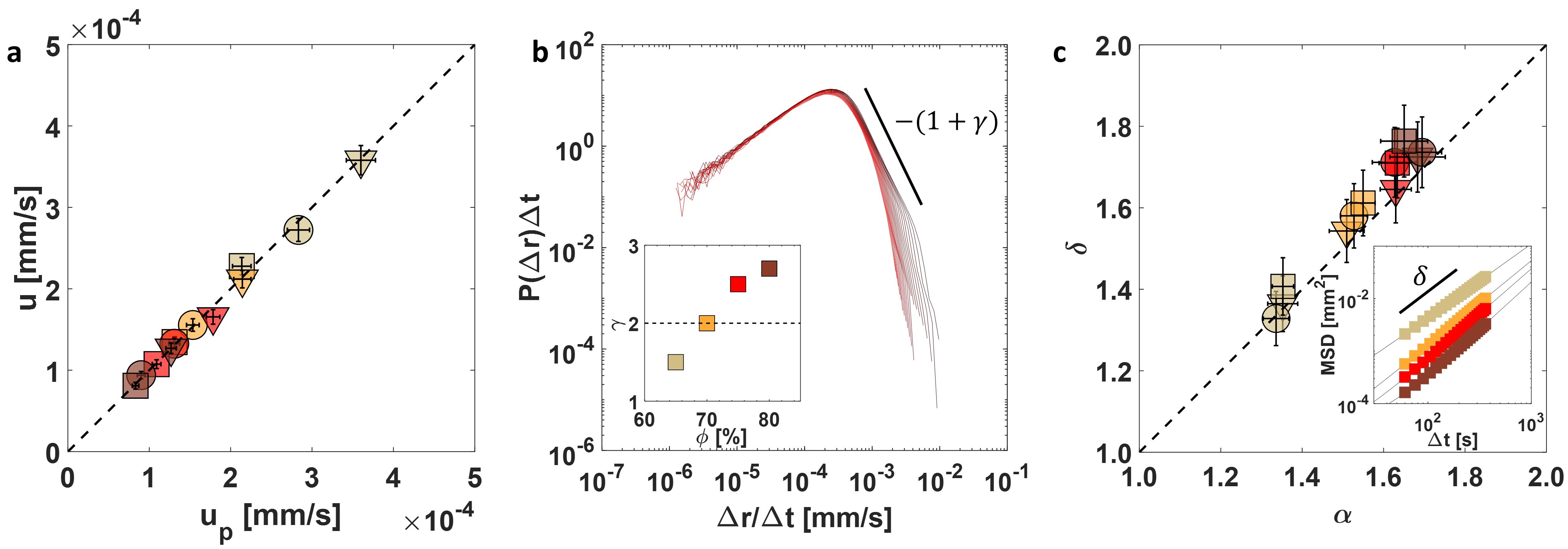}
    \caption{\textbf{Real-space bubble displacements}.
(a) Comparison between the characteristic velocity $u$ obtained with DDM and the velocity of the maximum peak of the bubble displacement distribution $u_p$ calculated with bubble tracking.
(b) Rescaled probability distribution functions of bubble displacements at different delays $\Delta t$ (color code going from black to red with increasing lag time) for the sample $\phi$ = 65\% at foam age $t^*$ = 3600 s.
The inset shows how the exponent $\gamma$ varies with $\phi$, exceeding the limit $\gamma$ = 2 marked by the horizontal dashed line.
(c) Comparison between the exponent $\delta$ measured from the power-law $\Delta t$-dependence of the MSD (inset from \cite{guidolin_tracking}) and the compressing exponent $\alpha$ measured from the intermediate scattering functions. The dashed line corresponds to the expected relation $\delta = \alpha$. Different colours correspond to different $\phi$, different symbols to different foam ages $t^*$, as indicated above.}
\label{fig:FIGURE_3}
\end{figure*}

Let us now compare the samples at fixed foam age.
The downward shift of $\Gamma(q)$ with increasing $\phi$ shown in Fig. \ref{fig:FIGURE_2}(b) reflects a decrease of $u$.
As the coarsening rate at a given foam age is essentially the same for all samples, the variation of $u$, and hence of $\Tilde{u}$, with $\phi$ can be entirely ascribed to the different rheology of the foam continuous phase.

A key observation to link continuous phase rheology to the observed dynamics is that the linear scaling of $\Gamma(q)$ is found at high values of $q$ corresponding to length scales much smaller than the typical bubble size.
Bubble displacements over such short length scales are too small to make the interstitial emulsion yield \cite{guidolin_tracking} and are therefore expected to correspond to elastic deformations of the foam matrix.
Motivated by this observation, we expect the high-$q$ bubble dynamics to be governed by continuous phase elasticity.
We thus plot the characteristic velocity $\Tilde{u}$ as a function of the emulsion elastic modulus $G_0$ in Fig. \ref{fig:FIGURE_2}(f).
In the same graph, we also plot the value of $\Tilde{u}$ calculated for shaving foam \cite{Giavazzi2021} as a horizontal bar.
Even though this value does not depend on the bubble size \cite{Giavazzi2021}, it could in principle depend on the foam liquid fraction.
However, we stress that shaving foam has a liquid fraction ($\sim$ 8\%) very close to the one of our samples ($\sim$ 10\%).
We can thus consider its value of $\Tilde{u}$ as a reference value for an equivalent aqueous foam having $G_0 \sim 0$.
From the graph, we can see that the normalised velocity for the sample with the lowest $G_0$ is very close to the one expected for an equivalent aqueous foam.
By contrast, a gradual reduction of $\Tilde{u}$ is observed with increasing emulsion elasticity.

In principle, the matrix elasticity is expected to affect foam coarsening once the elastic modulus $G_0$ overcomes the bubble capillary pressure $\gamma/R$ (with $\gamma$ the matrix surface tension and $R$ the typical bubble size), namely when the so-called elastocapillary number \cite{kogan2013mixtures,gorlier2017coupled}  $Ca_{el}=G_0 R/ \gamma$ becomes larger than one.\cite{Note4}
We can therefore estimate a critical elastic modulus $G_0^*$ corresponding to $Ca_{el}\simeq 1$, above which we expect continuous phase elasticity to start impacting the foam relaxation dynamics.
Using $\gamma \simeq$ 30 mN/m and an average bubble radius $R \simeq$ 0.44 mm, we find $G_0^* \simeq$ 68 Pa, which is highlighted in Fig. \ref{fig:FIGURE_2}(f) as a vertical dashed line.
We can see that our data consistently show a significant deviation from the aqueous foam behavior only at $G_0$ above $G_0^*$, while the sample at $\phi =$ 65\%, having $G_0 < G_0^*$, behaves akin to an aqueous foam.
\cite{Note5}

\subsection*{Direct space analysis corroborates DDM results}

We remark that the observation of a ballistic-like scaling of the relaxation rate is \textit{a priori} compatible with different microscopic scenarios, including the presence of a spatially correlated flow \cite{drechsler2017active} or even a global drift \cite{castellini2024modeling}, as well as the independent persistent motion of single particles \cite{wilson2011differential}.
However, real-space bubble tracking in our systems has already pointed out the presence of persistent bubble motion over short length scales \cite{guidolin_tracking}. We hence exploit bubble trajectories to better understand the real-space meaning of the characteristic velocity $u$.
We start checking whether the mobility of individual bubbles depends on their radius $R$.
Only a mild dependence of the single-bubble mean square displacement on $R$ is observed in our systems (Supplementary Information, Fig. S9(b)), not compatible with the scaling $MSD\propto R^{-1}$ recently reported for a dense ripening emulsion \cite{rodriguez2023exp}.
Moreover, the spatial correlation properties of the displacement field are surprisingly similar between the samples and do not suggest the presence of long-range coordination in the bubble motion (Supplementary Information, Fig. S9(c)).

At each foam age $t^*$, the probability distribution of bubble displacements is found to systematically shift to larger displacements with increasing time delays $\Delta t$ at both low and high $\phi$ \cite{guidolin_tracking}.
The position of the distribution peak grows linearly over time, with a velocity $u_p$ matching the characteristic velocity $u$ obtained from DDM analysis, as shown in Fig. \ref{fig:FIGURE_3}(a). The observed distribution shift is thus fully consistent with persistent bubble motion with a typical velocity $u$.
Indeed, a simple normalization of both $x$ and $y$ axes with $\Delta t$ leads to an excellent data collapse, as shown in Fig. \ref{fig:FIGURE_3}(b) for $\phi$ = 65\% at $t^*$ = 3600 s.
Substantially equivalent results obtained at different foam ages and different oil fractions are reported in Supplementary Information (Fig. S5-S7).

This normalization highlights a power-law decay of the distributions at bubble displacements right above the maximum, before sharply dropping at the cut-off.
The probability density function is observed to decay as $P(\Delta r,\Delta t)\Delta t \sim (\Delta r/\Delta t)^{-(\gamma+1)}$, with exponent $\gamma$ increasing with $\phi$, as shown in Fig. \ref{fig:FIGURE_3}(b).

We recall that compressed exponential relaxations of the intermediate scattering function imply the presence of power-law tails in the probability density function of particle displacements.
Indeed, intermediate scattering functions decaying as $f(q,\Delta t) = e^{-(\Gamma(q)\Delta t)^\alpha}$, with $\Gamma(q) \sim q$, imply the probability density function of displacements to decay as $P(\Delta r,\Delta t) \sim \Delta r^{-(\alpha + 1)}$ at large $\Delta r$.
This is because the Fourier transform of a compressed exponential function is the Levy stable distribution which displays a power-law tail for large values of its argument.

However, the values of $\gamma$ obtained for $\phi$>65\% are well above the expected value $\gamma = \alpha$ and even exceed the asymptotic limit of Gaussian distribution $\gamma$ = 2.
This apparent discrepancy can be safely ascribed to the finite size of the sample. \cite{weron2001levy}.
Indeed, a significant overestimation of the tail exponent in finite samples can arise for compressing exponents $\alpha$ above $\simeq$ 1.5, as the Levy stable distribution develops a first steeper decay right past the maximum peak, before reaching the asymptotic decay with the correct exponent only at larger arguments.\cite{weron2001levy}
The experimental exponents $\gamma$ are in good agreement with the values calculated for the apparent steeper decay expected for similar compressing exponent $\alpha$ (Supplementary Information, Fig. S8).
The observation of the exact asymptotic power-law tail at high $\phi$ in our case is prevented by the presence of the cut-off length that physically limits the maximum bubble displacement.

If we now consider the bubble mean square displacement (MSD), defined as $\langle \Delta r^2\rangle = \int^{+\infty}_{0} r^2 P(r,\Delta t) dr$, when the compressing exponent $\alpha$ is smaller than 2, the MSD is infinite for every $\Delta t$.
However, the presence of the physical cut-off length avoids this unphysical situation: with such cut-off, the MSD is expected to scale as a power law $\sim \Delta t^\delta$, with exponent $\delta$ equal to the compressing exponent $\alpha$. \cite{Giavazzi2021} Experimentally, a scaling MSD $\sim \Delta t ^ \delta$ is observed for each sample, with exponent $\delta$ increasing with $\phi$ and gradually approaching the value 2 compatible with ballistic-like bubble motion \cite{guidolin_tracking}.
As shown in Fig. \ref{fig:FIGURE_3}(c), the exponent $\delta$ is in good agreement with the compressing exponent $\alpha$ obtained from the fit of the intermediate scattering function.

Overall, we find that changing the continuous phase rheology strongly modifies the structural relaxation of coarsening foams. Ballistic-like bubble displacements at small length scales are observed in samples with very different matrix rigidity. Increasing emulsion elasticity causes a reduction of the characteristic velocity associated to this bubble motion, as consistently observed in both direct and reciprocal space. Qualitative differences arise at larger bubble displacements, with emulsion plasticity governing the maximum extent of persistent displacements.
Increasing emulsion yield stress hinders mutual bubble displacements, affecting the overall shape of the bubble displacement distributions.
This results in a steeper scaling of the mean square displacement with increasing matrix stiffness, which translates into larger compressing exponents in the reciprocal space.

\section{Conclusions}
In this work, we investigated the internal dynamics associated with pressure-driven coarsening in foams made of dense emulsions.
The use of differential dynamic microscopy allowed an unprecedented characterisation of bubble dynamics in the reciprocal space for such complex foams with simple imaging experiments.
Like many other disordered non-equilibrium systems \cite{cipelletti2003universal}, foamed emulsions exhibit compressed exponential relaxations, associated with a ballistic-like linear dependency of the relaxation rate on the scattering wavevector.

We showed that changing the continuous phase rheology affects the shape of the intermediate scattering function, with stiffer emulsions yielding more compressed exponential decays.
Our work thus unveils the possibility to change the foam relaxation features by just modifying the rheological properties of the material between the bubbles.

DDM results are also in excellent agreement with real space observations obtained with bubble tracking, providing a robust link between direct and reciprocal space and significantly contributing to the understanding of internal stress-driven bubble dynamics.

As in coarsening aqueous foams \cite{Giavazzi2021}, the linear scaling of the foam relaxation rate is lost at a critical wavevector corresponding to the bubble length scale: the occurrence of bubble rearrangements within the foam suddenly changes the local stress configuration causing a loss of directional-persistency in the bubble motion.

Moreover, we showed that the characteristic velocity associated with the high-$q$ linear scaling of the relaxation rate corresponds to the typical velocity of bubble displacements in real space.
Once normalised with the coarsening rate, this velocity does not exhibit a significant variation within the range of bubble radii investigated, as already pointed out for shaving foam. \cite{Giavazzi2021}
However, we stress that while shaving foam reaches a self-similar growth regime after a transient time of the order of tens of minutes \cite{Durian1991_Gillette}, our foams are not expected to head towards a scaling state and thus we do not expect this velocity to remain constant indefinitely.
In the long run, foamed emulsions indeed deviate from traditional coarsening eventually breaking Plateau laws \cite{Guidolin2023}.
On the other hand, here we focused on the early stage of coarsening where possible deviations in the local structure are not enough to impact the bubble size distribution, which is similar at different $\phi$ and still resembles the one of traditional aqueous foams \cite{guidolin_tracking}.
The global foam structure can hence be considered in first approximation the same, enabling comparisons between the samples.

Furthermore, we shed a new light on the link between bubble dynamics and the rheological properties of the foam continuous phase.
We showed that while the characteristic velocity of persistent bubble motion at short length scales is governed by the elastic properties of the interstitial emulsion, emulsion plasticity starts playing a role only at larger length scales.
The gradual suppression of mutual bubble rearrangements with increasing emulsion yield stress \cite{guidolin_tracking} modifies the slope of the relaxation rate in the low-$q$ regime, towards the recovery of a fully linear dispersion relation.
This would suggest a limiting case where bubbles keep displacing without rearranging, deforming the foam and leading to unconventional bubble shapes. However, the existence of a finite emulsion yield stress does not completely impede bubbles from eventually rearranging, preventing the observation of a perfectly linear scaling over the whole range of accessible wavevectors $q$.

Beyond the new results on the structural relaxation of such complex foams, our experimental approach also shows how a tracking-free technique like DDM can be successfully used to extract in-depth information on the sample dynamics with simple imaging experiments.
This can be of particular interest for foam studies in general, as high-quality bubble segmentation is often impeded by the sample turbidity, as well as for studies of soft glassy materials composed of very small elementary constituents, which are typically studied with light, x-ray and neutron scattering in reciprocal space.

Finally, our work can pave the way for future experiments changing the nature of the foam continuous phase.
Extending to other systems would be of great relevance to establish a thorough link between matrix rheology and coarsening-related bubble dynamics.

\section*{Author Contributions}

All authors were involved in the conceptualisation of the experimental work.
C.G. performed the experiments.
F.G. and R.C. designed the methodology.
C.G. analysed the data and wrote the original draft. A.S. and F.G. supervised the project.
All authors discussed
the results and contributed to the final manuscript.

\section*{Conflicts of interest}
There are no conflicts to declare.

\section*{Acknowledgements}
We acknowledge illuminating discussions with Véronique Trappe.

\section{SUPPLEMENTAL MATERIAL}

\renewcommand\thefigure{S\arabic{figure}}
\setcounter{figure}{0}

\subsection{Extended DDM results}

\begin{figure*}[htbp]
    \centering
    \includegraphics[width=\textwidth]{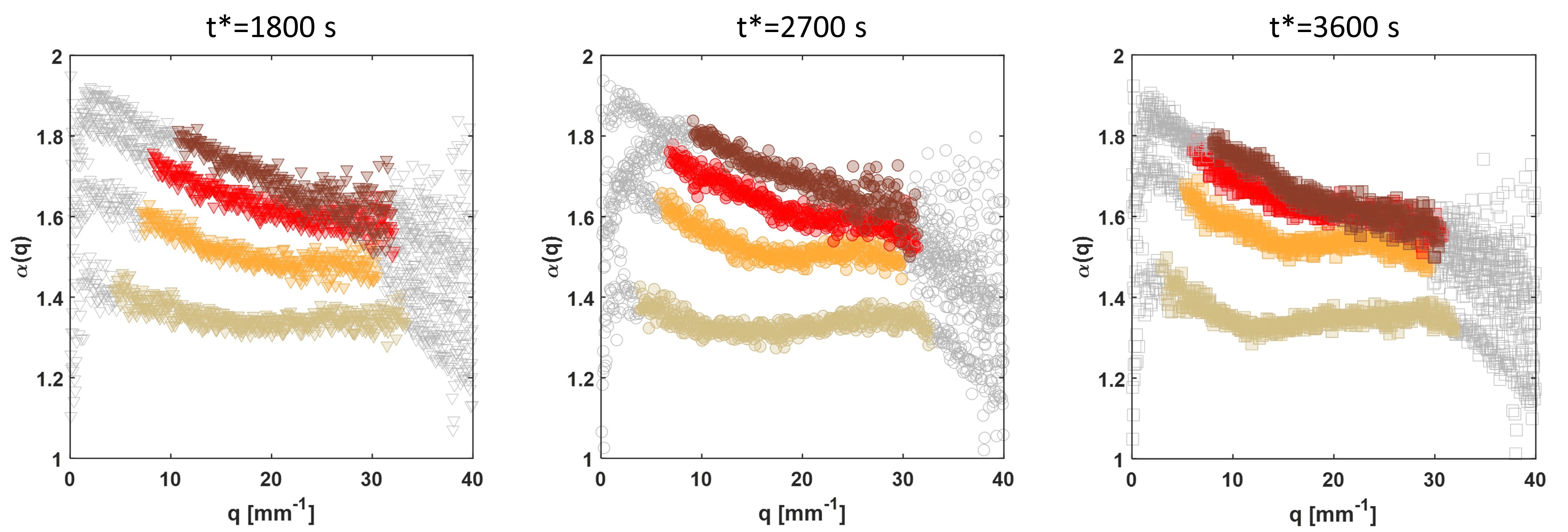}
    \caption{Compressing exponent $\alpha(q)$ obtained from a first fit of the image structure functions.}    \label{fig:SI_alpha_FreeFit}
\end{figure*}

The image structure function $d(q, \Delta t)$ obtained from DDM analysis is first fitted with a function of the kind $d(q,\Delta t)=A(q)[1-f(q,\Delta t)]+B(q)$, where $f(q,\Delta t)$ is a compressed exponential function $f(q,\Delta t)=\exp[-(\Gamma(q) \Delta t)^{\alpha(q)}]$ and the compressing exponent $\alpha(q)$ is left as a free parameter to check its dependence on $q$.
The compressing exponent $\alpha(q)$ is shown in Fig. \ref{fig:SI_alpha_FreeFit} for each $\phi$ at each foam age $t^*$, where we can see that it displays only a weak $q$-dependence.
We thus evaluate the average compressing exponent $\alpha = \langle \alpha(q) \rangle$ by averaging over a range of $q$ for which at least the first 10\% of the decay of $f(q,\Delta t)$ is visible and the amplitude is larger than the noise.

\begin{figure*}[htbp]
    \centering
    \includegraphics[width=\textwidth]{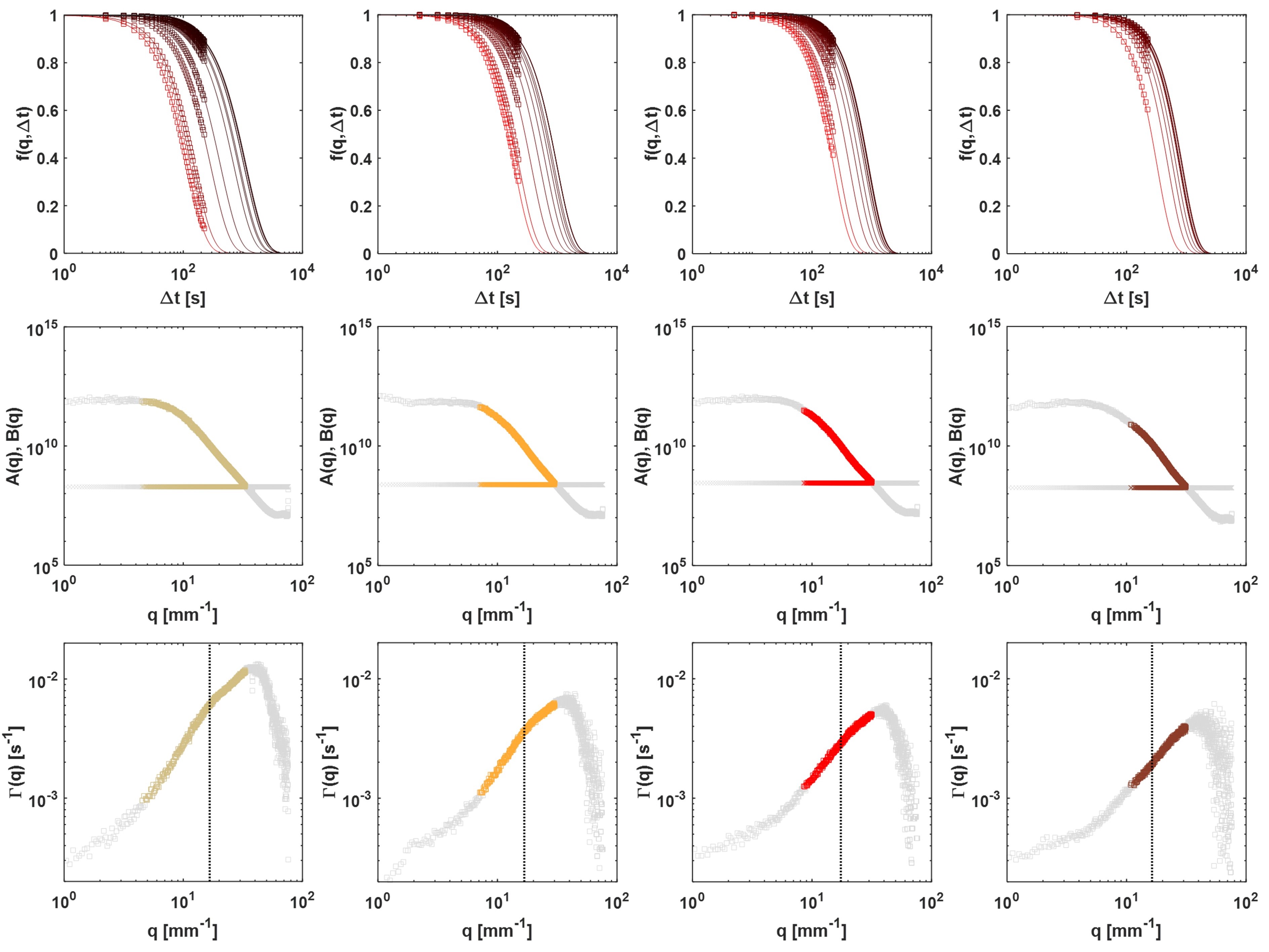}
    \caption{Fit results for the time window centered at $t^*=$ 1800 s.
    Each column correspond to a different oil fraction $\phi$ (65\%, 70\%, 75\%, and 80\% from left to right).
    First row: representative examples of intermediate scattering function $f(q,\Delta t)$ for different wavevectors $q$. Empty squares represent experimental data. The solid lines represent a compressed exponential fit to the data with fixed compressing exponent $\alpha$.
    Second row: amplitude $A(q)$ and noise term $B(q)$ obtained from the fit over the whole range of accessible $q$. 
    Coloured symbols highlight the selected range of $q$.
    Third row: relaxation rate $\Gamma(q)$. The vertical dashed line marks the wavevector $q^* = 2\pi/R^*$.}  \label{fig:SI_FitFixedExp_TW1800}
\end{figure*}

\begin{figure*}[htbp]
    \centering
    \includegraphics[width=\textwidth]{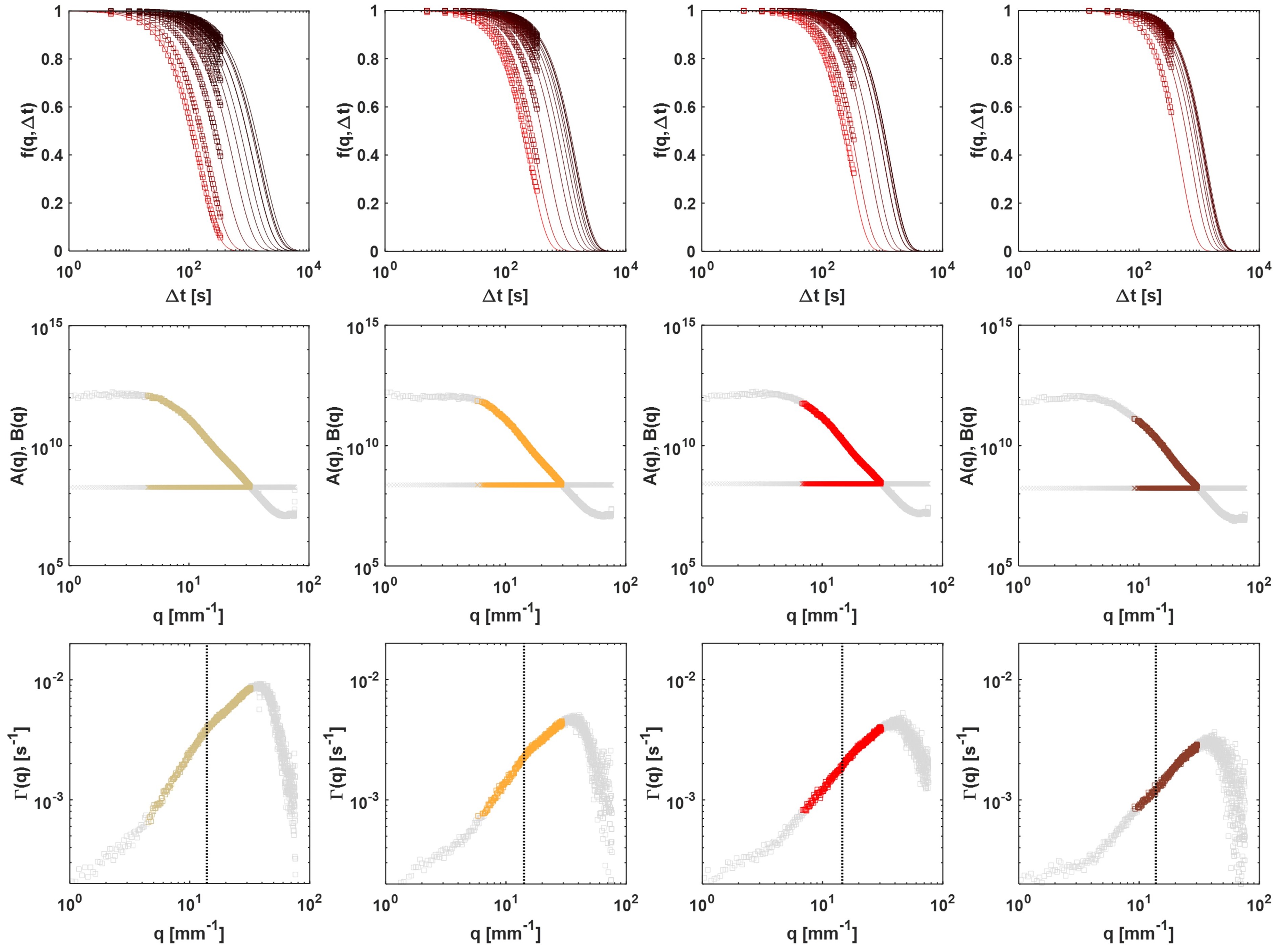}
    \caption{Fit results for the time window centered at $t^*=$ 2700 s.
    Each column correspond to a different oil fraction $\phi$ (65\%, 70\%, 75\%, and 80\% from left to right).
    First row: representative examples of intermediate scattering function $f(q,\Delta t)$ for different wavevectors $q$. Empty squares represent experimental data. The solid lines represent a compressed exponential fit to the data with fixed compressing exponent $\alpha$.
    Second row: amplitude $A(q)$ and noise term $B(q)$ obtained from the fit over the whole range of accessible $q$. Coloured symbols highlight the selected range of $q$.
    Third row: relaxation rate $\Gamma(q)$. The vertical dashed line marks the wavevector $q^* = 2\pi/R^*$.}   \label{fig:SI_FitFixedExp_TW2700}
\end{figure*}

\begin{figure*}[htbp]
    \centering
    \includegraphics[width=\textwidth]{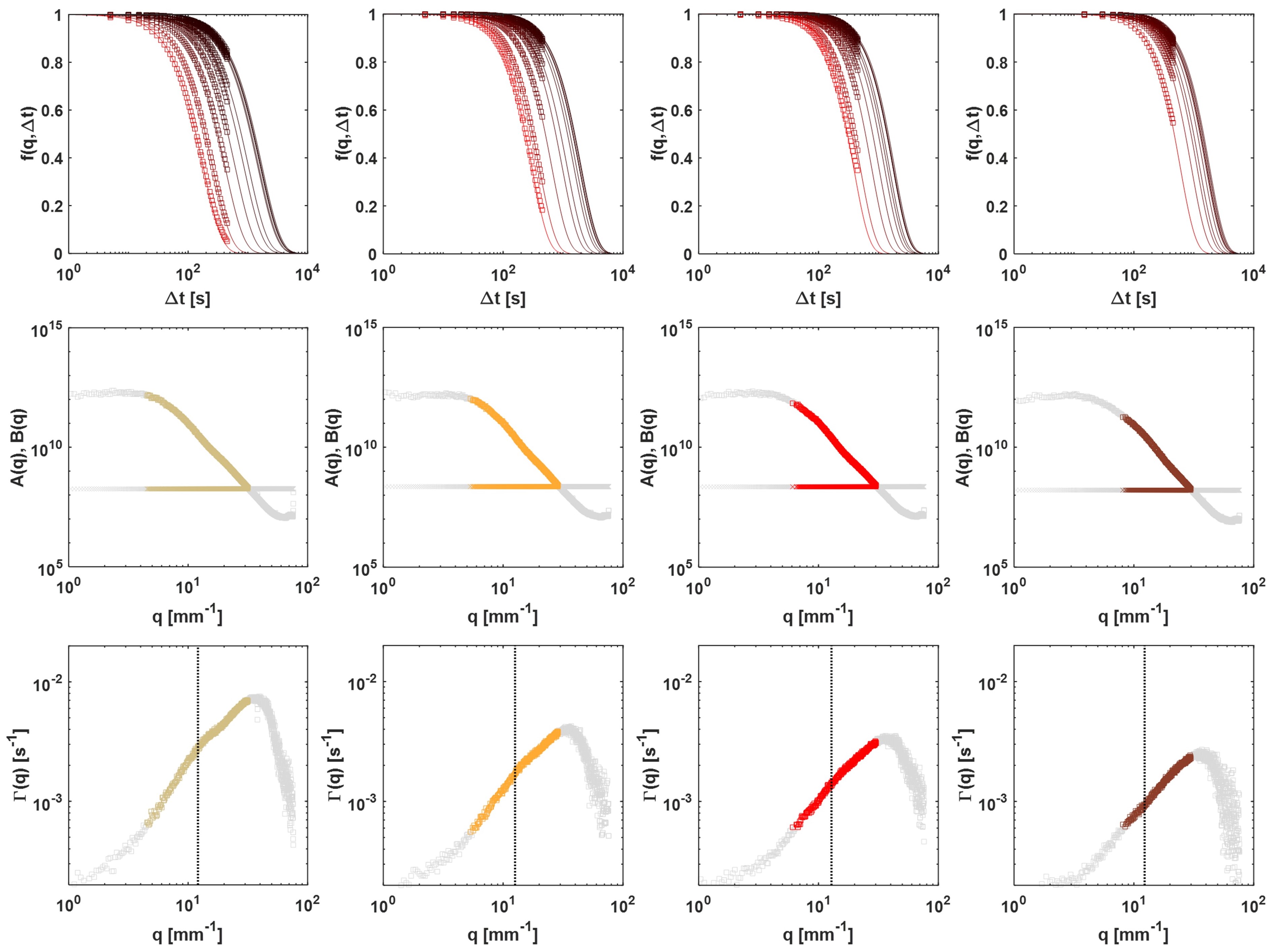}
    \caption{Fit results for the time window centered at $t^*=$ 3600 s.
    Each column correspond to a different oil fraction $\phi$ (65\%, 70\%, 75\%, and 80\% from left to right).
    First row: representative examples of intermediate scattering function $f(q,\Delta t)$ for different wavevectors $q$. Empty squares represent experimental data. The solid lines represent a compressed exponential fit to the data with fixed compressing exponent $\alpha$.
    Second row: amplitude $A(q)$ and noise term $B(q)$ obtained from the fit over the whole range of accessible $q$. Coloured symbols highlight the selected range of $q$.
    Third row: relaxation rate $\Gamma(q)$. The vertical dashed line marks the wavevector $q^* = 2\pi/R^*$.}    \label{fig:SI_FitFixedExp_TW3600}
\end{figure*}

We then fit the same data again with a compressed exponential where the compressing exponent is fixed to be equal to its mean value.
Results obtained from this fit are reported in Fig. \ref{fig:SI_FitFixedExp_TW1800}, \ref{fig:SI_FitFixedExp_TW2700}, and \ref{fig:SI_FitFixedExp_TW3600} for each sample at foam age $t^* =$ 1800, 2700, and 3600 seconds respectively.
Each figure contains representative examples of intermediate scattering functions $f(q,\Delta t)$ computed at different wavevectors $q$, the amplitude $A(q)$ and the noise term $B(q)$, as well as the relaxation rate $\Gamma(q)$.
The vertical dashed line in the graphs of $\Gamma(q)$ represents the wavevector corresponding to the a length scale of the order of the typical bubble size $R^*$, namely $q^* = 2\pi/R^*$. 
Each column refers to a different oil fraction $\phi$ (65\%, 70\%, 75\%, and 80\% from left to right).
The selected range of wavevectors $q$ is highlighted with coloured symbols.

\subsection{Distributions of bubble displacements}

\begin{figure*}[htbp]
    \centering
    \includegraphics[width=\textwidth]{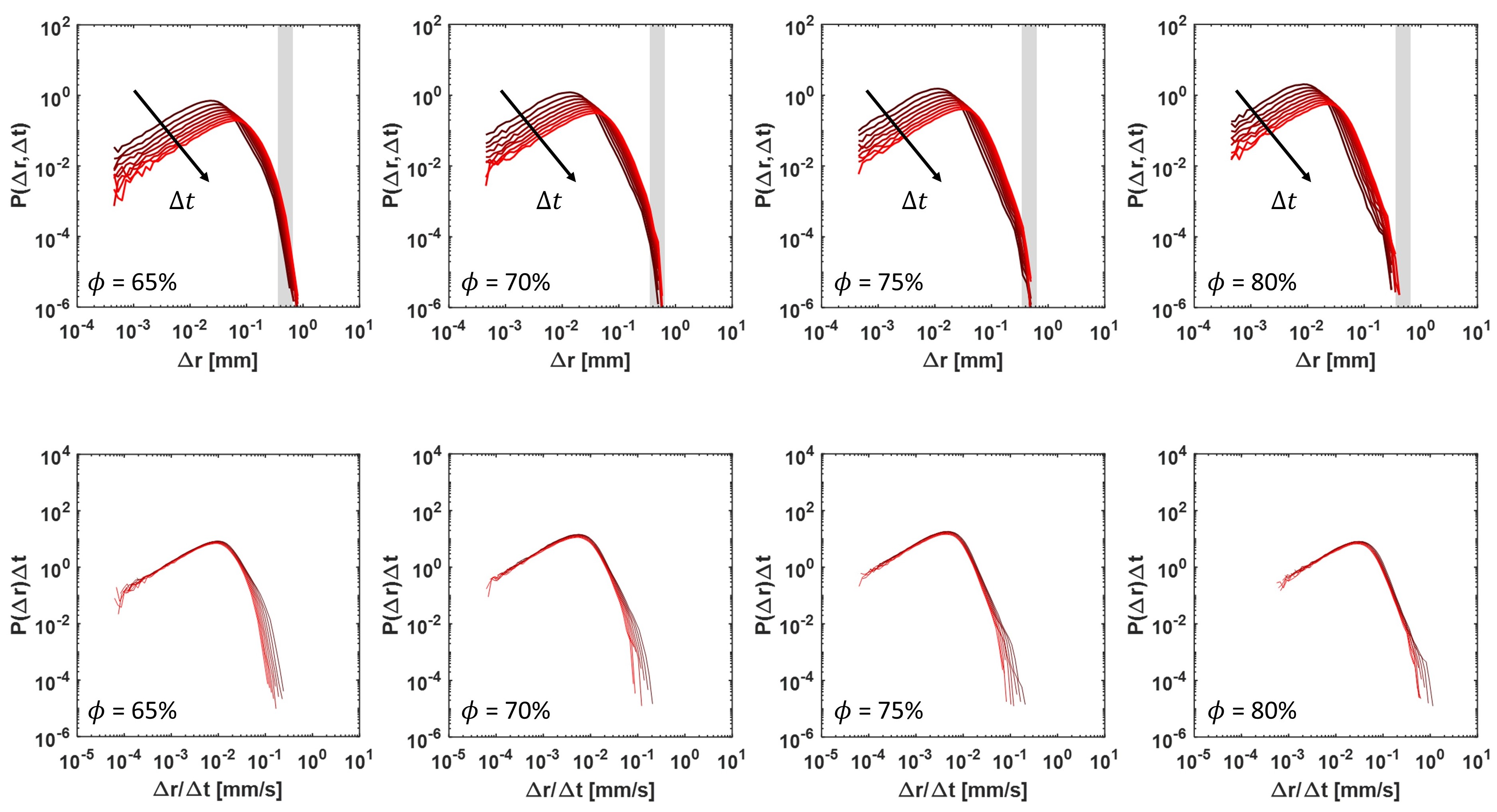}
    \caption{Distributions of bubble displacements for the time window centered around $t^*$ = 1800 s. Top row: probability distribution of bubble displacements $\Delta r$ at different time delays $\Delta t$. The vertical gray bar represents the typical bubble size. Bottom row: same distributions after normalisation with the time delay.}    \label{fig:SI_Distributions_TW1800}
\end{figure*}

\begin{figure*}[htbp]
    \centering
    \includegraphics[width=\textwidth]{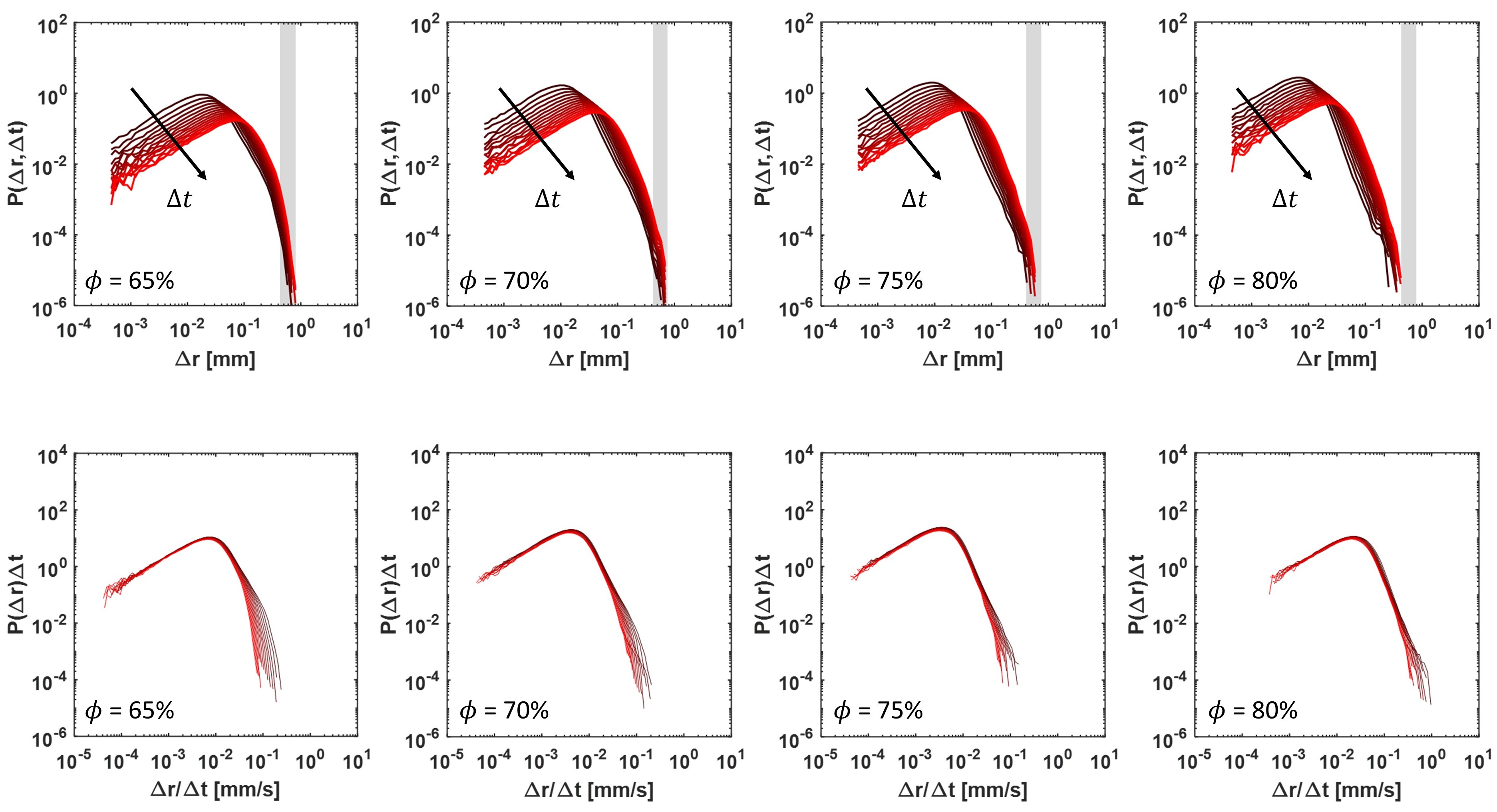}
    \caption{Distributions of bubble displacements for the time window centered around $t^*$ = 2700 s. Top row: probability distribution of bubble displacements $\Delta r$ at different time delays $\Delta t$. The vertical gray bar represents the typical bubble size. Bottom row: same distributions after normalisation with the time delay.}    \label{fig:SI_Distributions_TW2700}
\end{figure*}

\begin{figure*}[htbp]
    \centering
    \includegraphics[width=\textwidth]{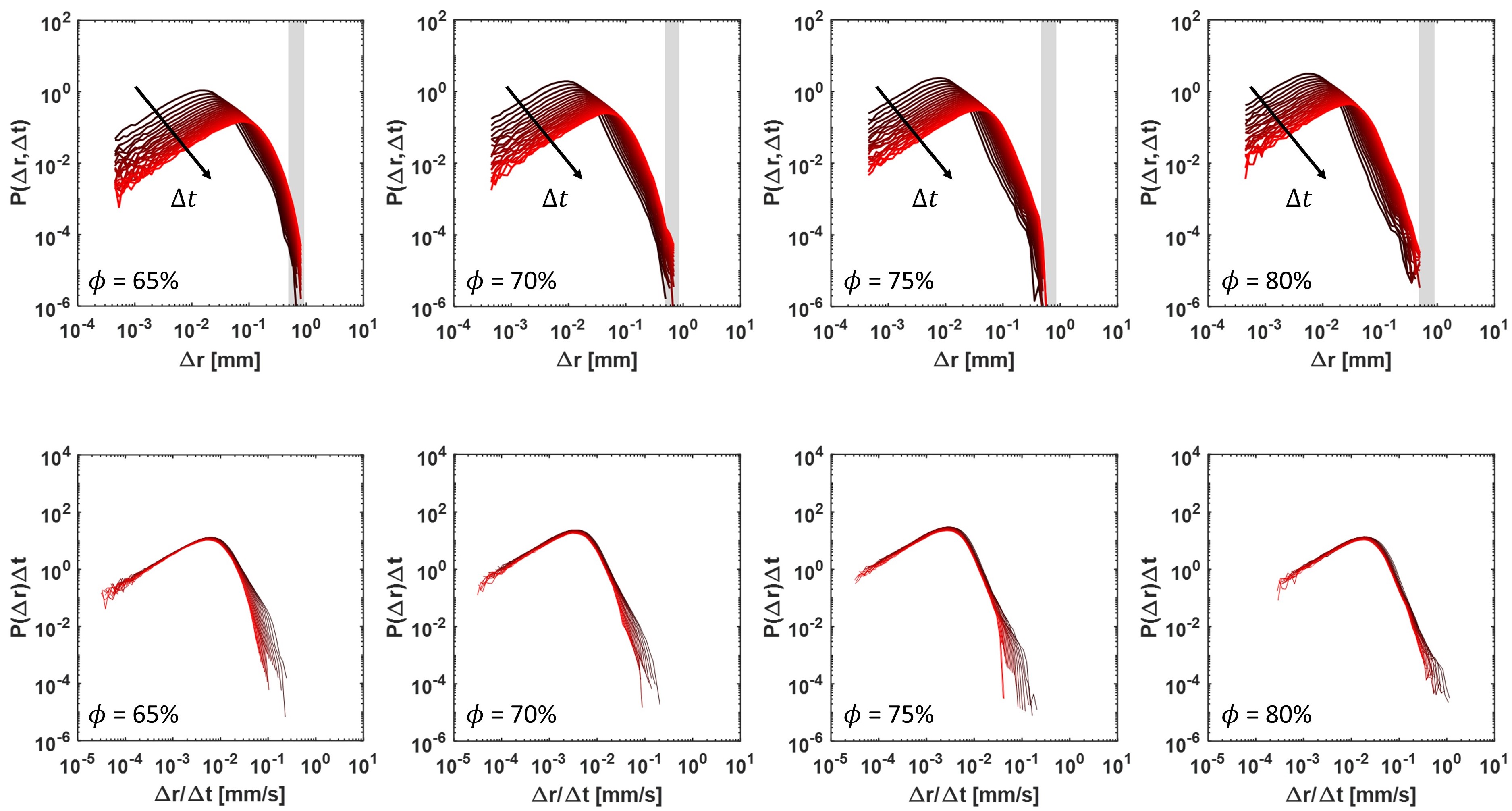}
    \caption{Distributions of bubble displacements for the time window centered around $t^*$ = 3600 s. Top row: probability distribution of bubble displacements $\Delta r$ at different time delays $\Delta t$. The vertical gray bar represents the typical bubble size. Bottom row: same distributions after normalisation with the time delay.}    \label{fig:SI_Distributions_TW3600}
\end{figure*}

For each foam sample, we calculated the distribution of bubble displacements $\Delta r$ for increasing time delays $\Delta t$ (from 60 to $t^*$/10 seconds with a step of 15 seconds) at foam age $t^*$ = 1800, 2700, and 3600 seconds, as reported in figures \ref{fig:SI_Distributions_TW1800}, \ref{fig:SI_Distributions_TW2700}, and \ref{fig:SI_Distributions_TW3600} respectively.

For each $\phi$ and $t^*$, the probability distribution of bubble displacements systematically shifts to larger displacements with increasing time delays $\Delta t$ \cite{guidolin_tracking}.
The position of the distribution peak grows linearly over time: a simple normalization with the lag time $\Delta t$ indeed leads to an excellent data collapse, consistent with ballistic-like bubble motion at short length scales.

This normalization highlights a power-law decay $P(\Delta r,\Delta t)\Delta t \sim (\Delta r/\Delta t)^{-(\gamma+1)}$, with exponent $\gamma$ increasing with $\phi$, at bubble displacements right above the maximum peak, before sharply dropping at displacements of the order of the typical bubble size.\\

\subsection{Apparently steeper power-law decay of pdf($\Delta r$) for large compressing exponents}

As discussed in the main text, if the intermediate scattering function is a compressed exponential of the kind $f(q,\Delta t)=\exp(-(v_0 q \Delta t)^{\alpha})$, the probability distribution of particle displacements pdf($\Delta r$) is expected to have a power-law tail decaying as $\sim -(1+\gamma)$ with $\gamma = \alpha$.

However, in foams at $\phi$ larger than 65\% we observe a steeper power-law decay of pdf($\Delta r$), with $\gamma > \alpha$.
This can be ascribed to the finite size of the sample \cite{weron2001levy} which, at large compressing exponents, results in an initially steeper decay of the probability distribution of displacements.

\begin{figure*}[htbp]
    \centering
    \includegraphics[width=\textwidth]{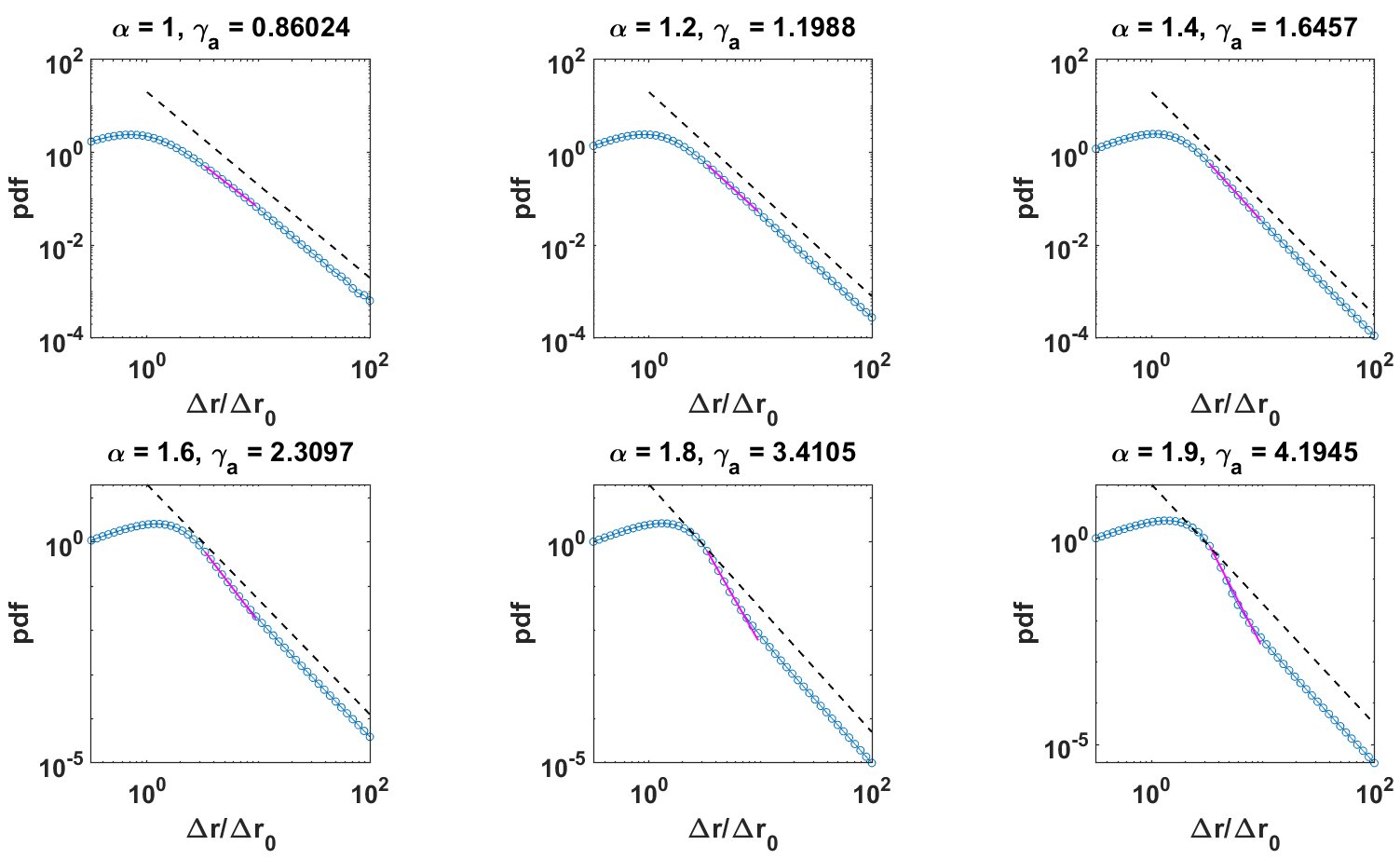}
    \caption{The graphs show the pdf of displacements expected for compressed exponential ISF having a compressing exponent $\alpha_t$.
    The dashed line marks the expected power-law decay $\sim -(1+\gamma)$ with $\gamma = \alpha$.
    The apparent steeper decay with exponent $-(1+\gamma_a)$ with $\gamma_a > \alpha$, obtained from a power law fit, is indicated by the magenta solid line.}    \label{fig:SI_pdf_simul}
\end{figure*}

In Fig. \ref{fig:SI_pdf_simul} we report the pdf($\Delta r$) calculated for compressing exponents $\alpha$ between 1 and 1.9.
We call $\gamma_a$ the apparent exponent measured immediately after the maximum peak.
We can see that, when $\alpha$ is only slightly larger than 1, we immediately recover the right scaling with $\gamma_a \simeq \alpha_t$.
For the sample $\phi=65\%$, indeed, we experimentally observe the right exponent, $\gamma=\alpha$.
However, when $\alpha$ is larger than 1.5, the distributions exhibit a steeper decay with exponent $-(1+\gamma_a)$, with $\gamma_a > \alpha$, before reaching the right scaling.
This explains why in our samples at higher $\phi$ we observe a power law tail steeper than expected.
We stress that in our case we could not reach the right scaling as the bubble displacements are restricted to length scales smaller than the typical bubble size.\\

\newpage
\subsection{Displacement correlations}

\begin{figure*}[htbp]
    \centering
    \includegraphics[width=\textwidth]{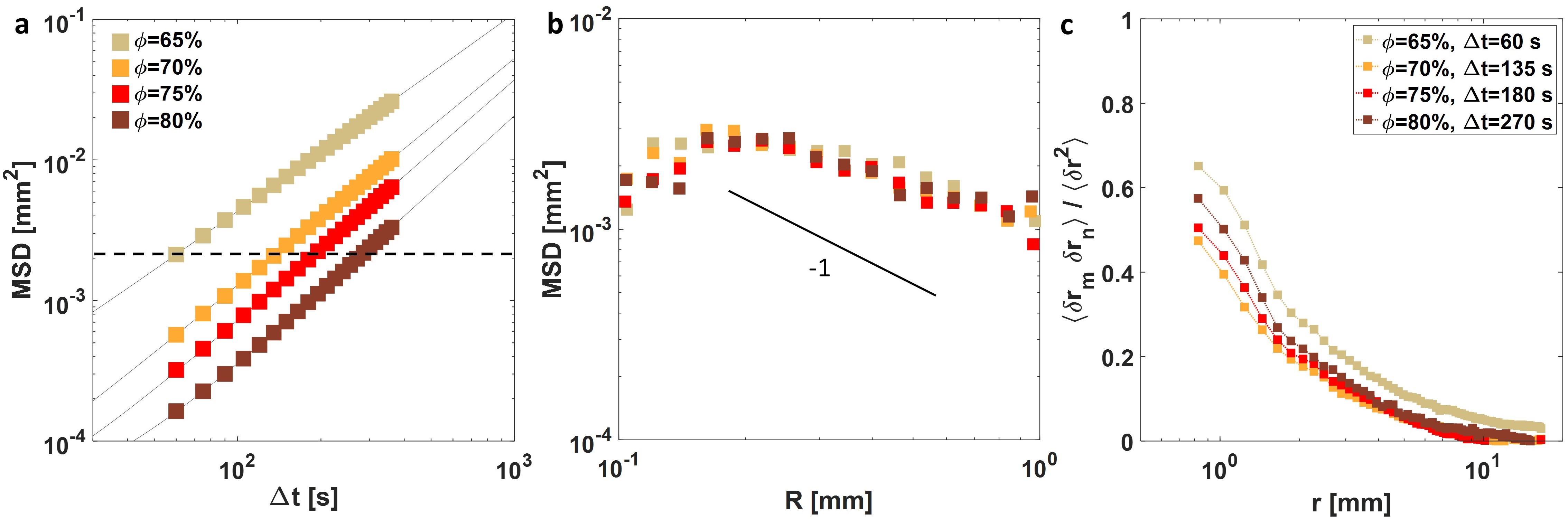}
    \caption{\textbf{Displacement correlations.} (a) $\Delta t$-dependence of the global MSD at different $\phi$ (from \cite{guidolin_tracking}). (b) Dependency of the single-bubble MSD on the bubble radius. The samples are compared at different time delays corresponding to the same global MSD, as indicated by the horizontal dashed line in (a).
    (c) Displacement correlation for the different samples evaluated at time delays corresponding to the same MSD.} \label{fig:SI_MSDvsR_CORR}
\end{figure*}

We investigate how single-bubble motility depends on the bubble size.
To account for the different average motility of different samples, we consider for each sample a different time delay $\Delta t$, corresponding to the same global mean square displacement (MSD), as shown in Fig. \ref{fig:SI_MSDvsR_CORR} (a).
Thus, we calculate the MSD for each single bubble and plot it against the bubble radius $R$.
Results are shown in Fig. \ref{fig:SI_MSDvsR_CORR}(b) for all oil fractions $\phi$. Only a mild dependency on the bubble size is observed, not compatible with the scaling $MSD\propto R^{-1}$ recently reported for a dense ripening emulsion \cite{rodriguez2023exp}.

For the same time delays, we also compute the spatial correlation function of bubble displacements as $\langle \delta \mathbf{r}_m \cdot \delta \mathbf{r}_n \rangle (r) / \langle \delta r^2 \rangle$, where $n$ and $m$ label distinct bubbles at a distance $r$, while $\delta \mathbf{r}_n$ and $\delta \mathbf{r}_m$ are the respective displacements.
Results are compared in Fig. \ref{fig:SI_MSDvsR_CORR}(c), where we can see that the correlation functions do not change significantly with $\phi$, revealing that the correlation properties of the displacement field are similar between all samples.\\

\end{document}